\documentclass[a4paper,11pt]{article}
\usepackage{jcappub}

\title{ Black holes inside cosmic voids}
\author[a]{Francisco Bento Lustosa,}
\author[b]{Milko Estrada,}
\author[c]{Marcony S. Cunha}
\author[a]{and Celio R. Muniz}
\affiliation[a]{Universidade Estadual do Ceará (UECE), Faculdade de Educação, Ciências e Letras de Iguatu, Av. Dário Rabelo s/n, Iguatu - CE, 63.500-000, Brazil.}
\affiliation[b]{Facultad de Ingeniería y Empresa, Universidad Católica Silva Henríquez, Chile.}
\affiliation[c]{Universidade Estadual do Ceará (UECE), Centro de Ciências e Tecnologia (CCT), 60714-903, Fortaleza-CE, Brazil.}
\emailAdd{chico.lustosa@uece.br}
\emailAdd{milko.estrada@gmail.com}
\emailAdd{marcony.cunha@uece.br}
\emailAdd{celio.muniz@uece.br}

\abstract{
This study examines the gravitational and thermodynamic properties of static, spherically symmetric black holes within cosmic voids -- vast underdense regions of the universe. By deriving a novel solution based on a universal density profile for voids, we analyze its spacetime structure, which reveals two horizons: One of the black hole and the other related to the de Sitter-like behavior. As the void approaches a perfect vacuum, the black hole horizon diminishes, tending to that of the Schwarzschild solution, while the outer horizon increases. We also study the solution stability via sound speed of the fluid, as well as the thermodynamic properties, including Hawking temperature, evaporation time, entropy, and specific heat. Our results show that as the void empties, the Hawking temperature rises, shortening evaporation times. The entropy follows the area's law and specific heat exhibits a minimum for a given black hole size, indicating a thermal transition and highlighting the role of voids in the black hole evolution. These findings offer new insights into the relationship between local gravitational collapse and large-scale cosmic structure, enhancing our understanding of the black hole behavior in underdense environments. We also provide a glimpse of a potential thermodynamic interaction between the event horizon and the cosmological horizon.
}
\begin{document}
\maketitle
\flushbottom
\section{Introduction}
Cosmic voids, the vast underdense regions that permeate the large-scale structure of the universe, have emerged as a critical area of study in modern cosmology \cite{Dubinski1993, Peebles2001, Sheth2004, Weygaert2011}. These regions, characterized by matter densities significantly lower than the cosmic average, occupy a substantial fraction of the universe's volume and play a pivotal role in shaping its evolution \cite{Bond1996, Alfaro2020}. The interplay between the formation of large scale filaments, expansion and merger of cosmic voids and the background evolution of our universe can provide explanations for some of the outstanding issues with our current standard cosmological model \cite{Hamaus_2014_1, Seifert2024}. The study of cosmic voids dates back to the early days of observational cosmology, when the first large-scale surveys revealed the ``bubble-like'' structure of the universe \cite{Kirshner1981}. Since then, advances in observational techniques, such as galaxy redshift surveys \cite{York2000, Colless2001, DESI2021}, have provided increasingly detailed maps of the cosmic web, highlighting the prevalence and importance of voids. However, recent cosmological studies suggest that cosmic voids may provide clues about the distribution of cosmic mass and serve as a useful tool to constrain cosmological parameters \cite{Hossen:2021etb,Contarini:2019qwf,Davies:2019yif}. These observations have been complemented by theoretical and numerical studies that have developed frameworks to understand the formation, evolution, and properties of voids \cite{Gomez-Vargas_2019, Rodriguez-Medrano2023}.

The gravitational potential of voids has significant implications for the dynamics of the universe. On large scales, voids contribute to the integrated Sachs-Wolfe effect, leaving imprints on the cosmic microwave background (CMB) \cite{Sachs1967, Planck2016}. On smaller scales, they influence the motion of galaxies and the growth of structure through processes such as void expansion and shell crossing \cite{Lavaux2008}. Moreover, the low-density environment of voids makes them natural laboratories for studying modified gravity theories, as the weak gravitational fields in these regions amplify deviations from General Relativity \cite{Hui2013, Clampitt2013}. In addition to their large-scale significance, cosmic voids have also been linked to compact objects, such as primordial\cite{Capozziello:2004sh,Fakhry:2022zum}, supermassive \cite{Constantin:2007ab, Habouzit:2019fij,Curtis:2024hnc}, and cosmological black holes \cite{Stornaiolo:2001nv}, as well as to traversable wormholes \cite{Ellis:2007dmt}. While black holes are typically associated with dense environments as galactic centers or clusters, their presence and dynamics within voids remain an underexplored area of research. Despite the extremely low matter density in cosmic voids, inhomogeneities can still arise, potentially leading to black hole formation. 

Understanding how the large-scale structure of our universe has come to be the way we observe today depends on explaining the origins and evolution of these underdense regions that originate in the early universe. It is clear that dark matter and black holes play a pivotal role in the formation and evolution of overdense regions \cite{Cunha:2022kep}, but the study of voids can offer a different type of observational window to test theoretical models \cite{Lester2021}. It has been suggested that primordial black holes could serve as seeds for cosmic voids and contribute to their stability during cosmological evolution \cite{Capozziello:2004sh}. It is also clear by now that a better understanding of the effects of voids in our current estimations of the rates of expansion of the universe is needed to clarify some claims made in the recent literature regarding the possibility of explaining the observed accelerated expansion with the timescape model of cosmological backreaction \cite{Seifert2024, Williams2024}.

It is worth mentioning that reference \cite{Habouzit:2019fij} argues that although low-mass galaxies are most common near the centers of cosmic voids, the specific star formation rate slightly increases in such locations, suggesting that stars can form efficiently in these voids. In this regard, reference \cite{Curtis:2024hnc} also argues that the specific star formation rates of galaxies in voids are higher than those of galaxies outside of voids. Additionally, it mentions that in the case of galaxies with a certain type of central supermassive black holes, the fraction of active galaxies in voids is approximately 25\% higher than that of active galaxies outside of voids. Based on their analysis, the authors of \cite{Habouzit:2019fij} argue that this phenomenon, related to the efficient formation of compact objects in cosmic voids, cannot only be attributed to the prevalence of low-mass galaxies, but also to the fact that the inner regions of the voids predominantly host low-mass black holes. In this regard, they found similarities between the simulations of black hole mass and galaxy mass. The authors suggest that even if the growth channels in cosmic voids differ from those in denser environments, voids may grow their galaxies and black holes in a similar way. Thus, it is intriguing to study the link between the nature of the matter distribution in voids (which, broadly speaking, could be associated with its energy density structure) and the formation of black holes. In this regard, reference \cite{Fakhry:2022zum} suggests the presence of dark matter halos in cosmic voids, implying that it is likely that Primordial Black Holes (PBHs) can be clustered in cosmic voids. In this context, the authors mention that the size of cosmic voids and their underdense nature make them suitable platforms for investigating primordial density fluctuations under which PBHs would have formed. 

One of the key aspects of void physics is their density profile, which describes how matter is distributed within and around these underdense regions. Empirical and theoretical models, such as the universal density profile proposed by \cite{Hamaus_2014_2}, have been instrumental in characterizing voids and their gravitational effects. These profiles often take the form $\rho(r)=\rho_0[1+\delta(r)]$, where $\rho_0$ is the averaged background density and $\delta(r)$ encodes the density contrast as a function of the radial distance from the void's center. The precise form of $\delta(r)$ depends on the void's size, environment, and the underlying cosmological model. For example, the profile proposed by \cite{Hamaus_2014_2} incorporates a compensation wall, where the underdense core is surrounded by an overdense shell, reflecting the gravitational interplay between voids and their surroundings.

In this work, we aim to bridge the gap between cosmic voids and black hole physics by exploring the gravitational and thermodynamic implications of universal void density profiles. By deriving a novel black hole solution, we seek to uncover new insights into the interplay between local gravitational collapse and global cosmological structure. This approach is built on previous studies of a void density profile studied in \cite{Hamaus_2014_2} and black hole thermodynamics \cite{Hawking1974, Bousso2002}, while extending the discussion to include the unique effects of underdense environments. As we will see, the black hole and the cosmic void establish a symbiotic relationship, where each contributes to the persistence of the other. The black hole can act as both a seed, at least the primordial ones \cite{Capozziello:2004sh}, and a source of stability for the void, while this latter, in turn, extends the black hole lifetime by increasing its evaporation time. This mutual influence suggests that voids not only aid in forming black holes but also help preserve them, reinforcing their coexistence within the cosmic structure. Through this analysis, we hope to contribute to a deeper understanding of the large-scale structure of the universe and the fundamental nature of gravity.

The paper is structured as follows. In Section II, we introduce our black hole solution, examining its structure and stability. In Section III, we explore its thermodynamics, with a particular focus on the Hawking temperature, evaporation time, and specific heat. Finally, in Section IV, we summarize our findings and conclude the paper. We will use $\hbar = c = G = k_B=1$.


\section{Voids' black hole solution}

The underdense regions of our universe are not empty and could host both Primordial Black Holes \cite{Fakhry:2022zum} and supermassive ones \cite{Constantin:2007ab, Curtis:2024hnc} inside active galactic nuclei (AGNs). Although the phenomenological impact of such scenarios has been explored in the context of cosmological models of large scale structure, we have not encountered works focused on the study of specific theoretical models taking into account the horizon structure of such black holes and their possible impact on the void internal dynamics. In this section, we derive analytically a black hole metric by assuming the existence of a central mass $M$ at the center of the void and integrating the matter distribution of the void as described in \cite{Hamaus_2014_2}.

\begin{figure}[h]
    \centering
    \includegraphics[width=0.6\textwidth]{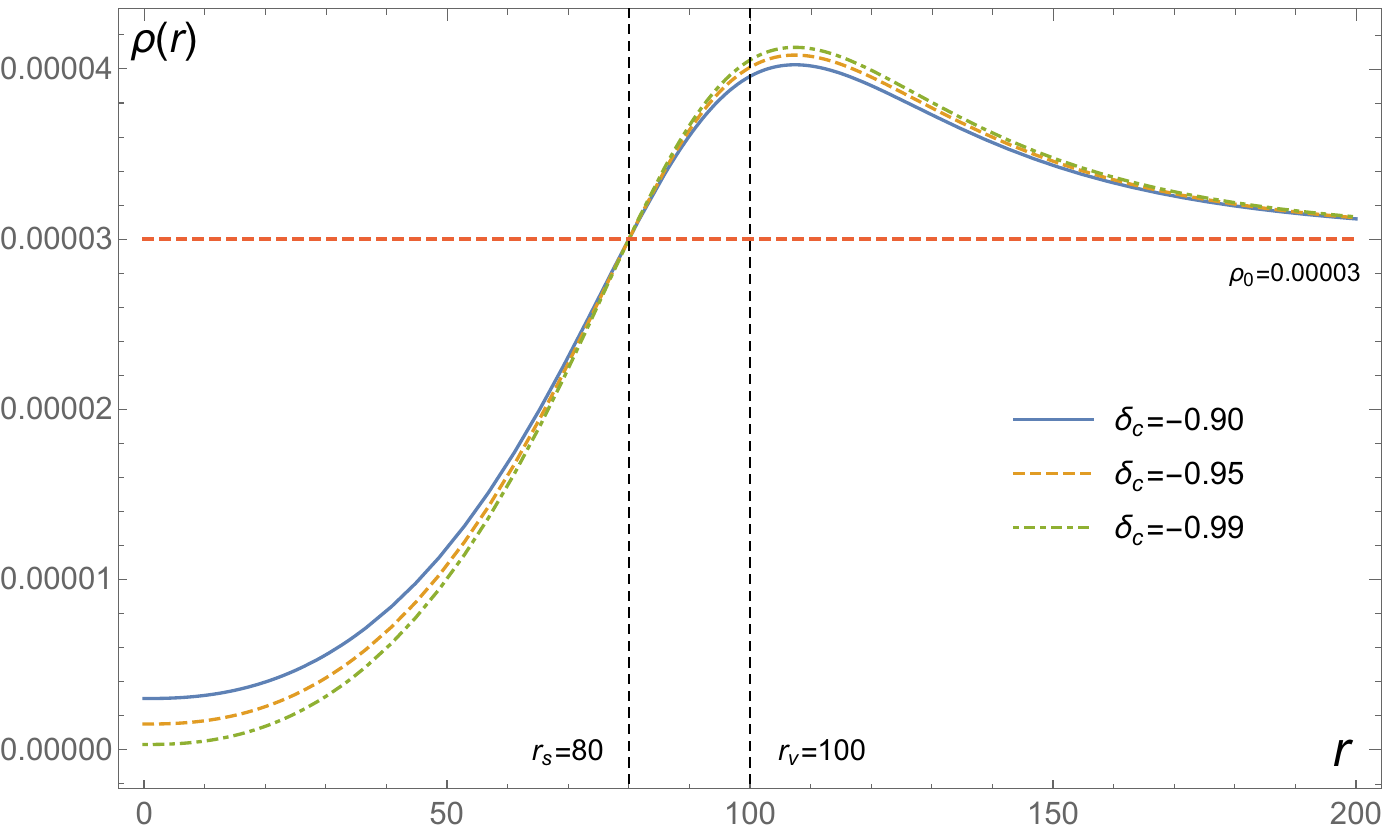} 
    \caption{Universal density profile $\rho(r)$ as a function of $r$, for some values of $\delta_c$, namely, $\delta_c=-0.90$ (solid blue), $\delta_c=-0.95$ (dashed orange), $\delta_c=-0.99$ (dotdashed green), considering the parameters $\rho_0=0.00003$, $\alpha=2.4$, $\beta=7.5$, $r_v=100$, and $r_s=80$.}
    \label{density}
\end{figure}

We derive the metric of the spacetime generated by the black hole inside he cosmic void from the  phenomenological universal density profile presented in \cite{Hamaus_2014_2}, given by
\begin{equation}\label{dens}
\rho(r)=\rho_0\left[1+\delta_c\frac{1-(r/r_s)^{\alpha}}{(1+(r/r_v)^{\beta}}\right],
\end{equation}
where $r_s$ represents the scale at which the density of the void equals the average of its surroundings, while $r_v$ denotes the void size. The parameter defined as $\delta_c$ (negative) quantifies the contrast, measuring the emptiness of the void, with $\delta_c=-1$ meaning a perfect vacuum at $r=0$, $\rho_0$ is the average density of the universe, and $\alpha$ and $\beta$ serve as empirical fitting parameters. In Figure \ref{density}, we represent the universal density profile for some values of $\delta_c$. Observe the formation of an overdense region, a ``wall'' around the edge of the void (for $r\geq r_s$), in which $\rho(r)>\rho_0$. Note also that the greater the absolute value of $\delta_c$, the deeper the void.

With this density profile as the material source for the black hole and considering an overdense initial central mass $M$ as an integration constant, we can find its mass function as being 
\begin{eqnarray} \label{FuncionDeMasa}
M(r)&=&\int 4\pi r^2 \rho(r) dr=M+\frac{4\pi}{3}r^3 \rho_0  
\!\!\left[1 + \delta_c \, {}_2F_1 \left(1, \frac{3}{\beta}, \frac{3 + \beta}{\beta}, -\!\left(\frac{r}{r_v}\right)^\beta \right)\qquad \qquad \right. \nonumber\\ 
&&\left. - \frac{3}{3 + \alpha} \left(\frac{r}{r_s}\right)^\alpha \delta_c \, {}_2F_1 \left(1, \frac{3 + \alpha}{\beta}, \frac{3 + \alpha + \beta}{\beta},\! -\left(\frac{r}{r_v}\right)^\beta \right)  \right], \qquad
\end{eqnarray}
where $_2F_1$ is the Gauss hypergeometric function. Thus, the metric coefficient of the black hole is 
\begin{equation} \label{FuncionF}
   f(r)=1-2 M(r)/r.
\end{equation}
The expansion near the origin for that coefficient is
\begin{equation}
    f(r)\approx 1-\frac{2 M}{r},
\end{equation}
{\it i.e.}, at this region the spacetime behaves as a Schwarzschild one. Therefore, the obtained black hole solution must be singular at the origin.

On the other hand, for larger values of the radial coordinate, such that $0 < r_s < r_v \ll r$, given the constraints of equations \eqref{alfa} and \eqref{beta}, which imply that $\beta > \alpha$, we observe that the energy density $\rho \to \rho_0$. Indicating that outside the void we recover the average distribution of matter of the universe, as expected. However, for a certain inner domain of the radial coordinate, an effective cosmological horizon arises inside the void, which represents an \textit{effective} boundary for the causal spacetime. In fact, considering now the region in which $0\ll r\ll r_s<r_v$, we have 
\begin{equation}\label{effdsitter}
    f(r)\approx 1-\frac{4}{3} \pi  \rho_0  (1+\delta_c)r^2.
\end{equation}
Thus, in this intermediate region, distant from both the origin and the void's edge, the spacetime of the black hole within a cosmic void closely resembles the de Sitter spacetime, therefore exerting a repulsive effect that counteracts the trend toward gravitational collapse. In this regime, the cosmological constant-like behavior of the void becomes dominant at large distances, driving a de Sitter-like expansion. Consequently, the black hole-cosmic void symbiosis provides a repulsive force that prevents the system from undergoing gravitational collapse.


Moreover, given that the ratio between the densities of dark energy ($\rho_{\Lambda}$) and total matter surrounding the void ($\rho_0$) is approximately 2.5 \cite{Planck:2015fie}, the effective cosmological constant within the void, $\Lambda_v= 8\pi \rho_0(1+\delta_c)$, is smaller than the universal one by the same factor. As the latter also permeates the void, these constants are summed and then the size of the almost empty structure hosting the black hole must increase faster than the overall cosmic expansion, which should be measurable. In fact, the expansion rate of cosmic voids has been explored in only a few inconclusive studies, with indications of a faster expansion \cite{Sato:1983ka,Galarraga-Espinosa:2023zmv}.

\begin{figure}[t]
    \centering
    \includegraphics[width=0.6\textwidth]{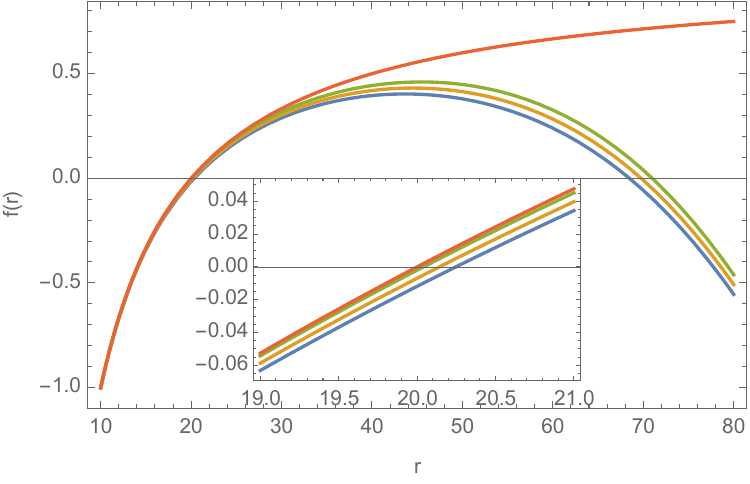} 
    \caption{Metric coefficient as a function of $r$, for some values of $\delta_c$, from bottom to top, namely, $\delta_c=-0.90$ (blue), $\delta_c=-0.95$ (orange), $\delta_c=-0.99$ (green), and Schwarzschild ($\rho_0=0$, red), considering the parameters $\rho_0=0.00003$, $\alpha=2.4$, $\beta=7.5$, $r_v=100$, $r_s=80$, $M=10$.}
    \label{fig1}
\end{figure}

In Fig. \ref{fig1}, we plot the metric coefficient as a function of the radial coordinate $r$, for some values of $\delta_c$, the contrast parameter that indicates how `empty' is the void. The plot indicates the presence of two horizons: one associated with the black hole and another horizon \textit{external}, which resembles the typical behavior of a cosmological horizon. In particular, the former increases in size, while the latter shrinks as the modulus of the contrast parameter $\delta_c$ decreases. Conversely, it can be viewed as the de Sitter-like horizon taking up larger regions of the void as it empties ($\delta_c\to -1$), which in turn can lead to a negative pressure in the region between the two horizons. This means that a void could eventually contain a type of de Sitter bubble. To understand how this new structure would interact with its surroundings would require a better understanding of void walls, which was exactly one of the motivations behind the new universal model of reference \cite{Hamaus_2014_2}, but this points to a possible connection with the timescape model, where voids play a pivotal role in driving the observed accelerated expansion of our universe \cite{Seifert2024, Williams2024}.


\subsection{Structure of the horizons}
First, we can note from equations \eqref{FuncionDeMasa} and \eqref{FuncionF} when $f(r)=0$ the mass parameter $M$ becomes a function given by: 
\begin{align}\label{BHmass}
 &M=   \frac{-r}{3 (3 + \alpha)}  \left\lbrace -9 + 12 \rho_0 \pi r^2 - 3 \alpha + 4 \rho_0 \pi r^2 \alpha + 12 \rho_0 \pi r^2 \delta_c \, _2F_1\!\left( 1, \frac{3}{\beta}, \frac{3 + \beta}{\beta}, -\left(\frac{r}{r_v}\right)^\beta \right) \right.\nonumber \\ 
& \left. + 4 \rho_0 \pi r^2 \delta_c \left[\alpha~_2F_1\!\left( 1, \frac{3}{\beta}, \frac{3 + \beta}{\beta}, -\left(\frac{r}{r_v}\right)^\beta \right)\!- 3 \left(\frac{r}{r_s}\right)^\alpha \!\! {}_2F_1\!\left( 1, \frac{3 + \alpha}{\beta}, \frac{3 + \alpha + \beta}{\beta}, -\left(\frac{r}{r_v}\right)^\beta\right) \right] \right\rbrace.
\end{align}

To constrain the parameters $\alpha$ and $\beta$ we go back to the phenomenological analysis of reference \cite{Hamaus_2014_2}, where it was demonstrated that appropriate values in the parameter space are given by:
\begin{equation} \label{alfa}
\alpha(r_s) = -2 \left( \frac{r_s}{r_v} - 2 \right), 
\end{equation}

\begin{equation} \label{beta}
\beta(r_s) =
\begin{cases}
17.5 \frac{r_s}{r_v} - 6.5 & \text{for } \frac{r_s}{r_v} < 0.91 \\
-9.8 \frac{r_s}{r_v} + 18.4 & \text{for } \frac{r_s}{r_v} > 0.91,
\end{cases}
\end{equation}
where it is useful to remember that $r_s$ has to do with the scale where the density of the void becomes equal to the average density of the surrounding universe, which has to do with the wall like structure that can form at the edge of the void \cite{Hamaus_2014_2}.

\begin{figure}[ht]
    \centering
    \includegraphics[width=0.6\textwidth]{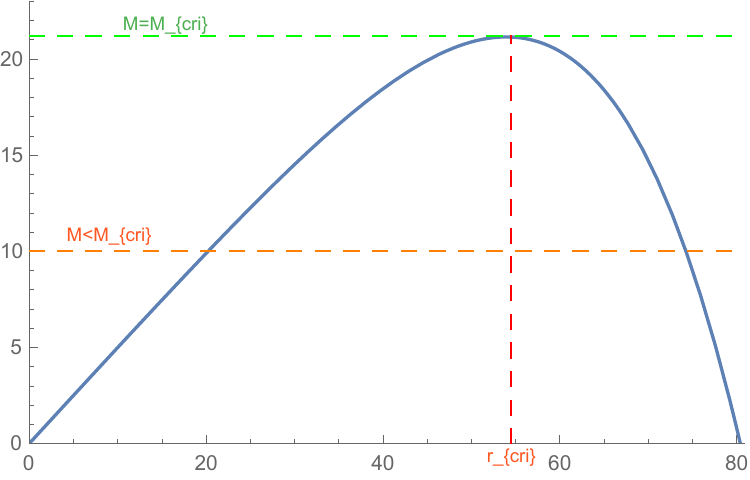} 
    \caption{On the horizontal axis $r$ such that $f(r) = 0$. On the vertical axis the Mass parameter $M$ for $r_s=80,r_v=100, \delta_c=-0.9,\rho_0=0.00003$}
    \label{figMasa1}
\end{figure}

Using these adjustments, in Figure \ref{figMasa1} we show the generic behavior of the mass parameter for the case where $r_s/r_v < 0.91$. We can verify that this behavior is generic for the second case of equation \eqref{beta}. The ascending curve on the left represents the values of the event horizon such that $f(r_h) = 0$. The descending curve on the right corresponds to the values of the outer horizon (analogous to the cosmological horizon) such that $f(r_{++}) = 0$. 

\begin{figure}[ht]
    \centering
    \includegraphics[width=0.8\textwidth]{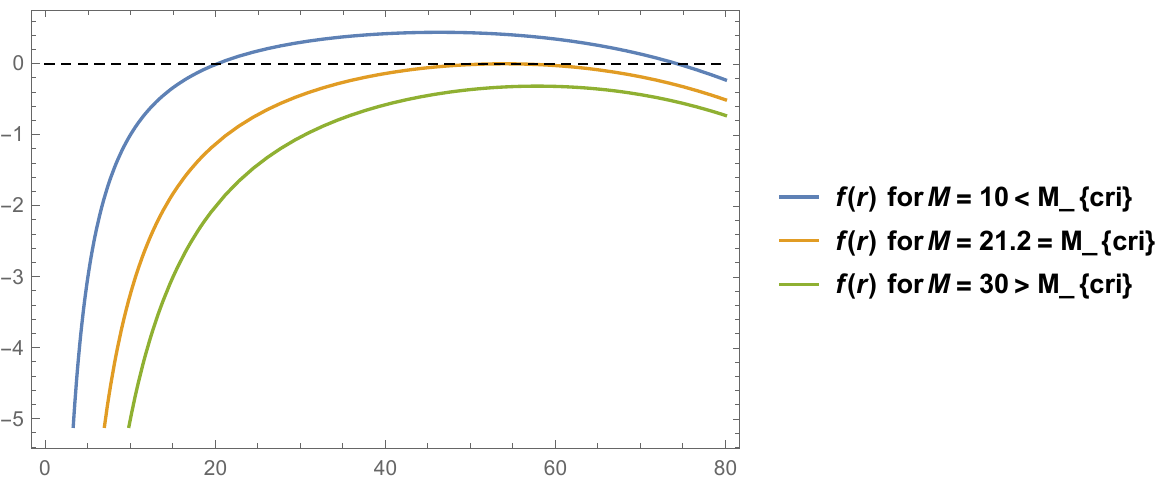} 
    \caption{Metric coefficient $f(r)$ as a function of r for $M<M_{crit}$ (upper blue curve), $M=M_{crit}$ (orange curve below), and $M>M_{crit}$ (lower green curve). We are using $r_s=80,r_v=100, \delta_c=-0.90$, and $\rho_0=0.00003$.}
    \label{figFuncionF1}
\end{figure}

We observe that the red vertical line corresponds to the value of the critical radius $r_{\text{cri}} = r_h = r_{++}$ (in figure $r_{\text{cri}} \sim 54.5$), where both the event horizon and the outer horizon coincide. Thus, for this value of the critical radius, we observe that the green horizontal line corresponds to a critical value $M = M_{\text{cri}}$, \textit{i.e.}, $M_{\text{cri}} = M(r_{\text{cri}} = r_h = r_{++})$ (in figure $M_{\text{cri}} \sim 21.2$), where both the event horizon and the outer horizon coincide. This situation is analogous to what occurs in the dS-Schwarzschild black hole solution when studying the thermodynamic equilibrium \cite{Ginsparg:1982rs}. The latter reference describes that the Nariai solution \cite{Nariai:1999iok} is obtained when the event horizon of the dS-Schwarzschild black hole approaches the cosmological horizon. On the other hand, for values lower than the critical mass value, such as the red horizontal line, we observe the presence of both horizons. Values higher than the critical mass value lead to naked singularities. This described situation can also be observed in Figure \ref{figFuncionF1}, which shows the behavior of the function $f(r)$.

As we shall see in the following discussions on the thermodynamic properties of our black hole solutions, different combinations of parameters can lead to very different physical pictures. On one side, we can consider the void as a large enough region of space-time and the distance between the two horizons to be sufficiently large so we can discuss the effects of the inner horizon separately and estimate its impact on the inner region of the void. Depending on its initial mass, the density of matter in the void and the parameters regulating its scale, one can have an evaporating black hole, and the outer horizon effectively vanishes with as it expands beyond the voids edge (where the overall cosmological background would have to be taken into account for a consistent description).

\subsection{Solution's stability}

\begin{figure}[t]
    \centering
    \includegraphics[width=0.6\textwidth]{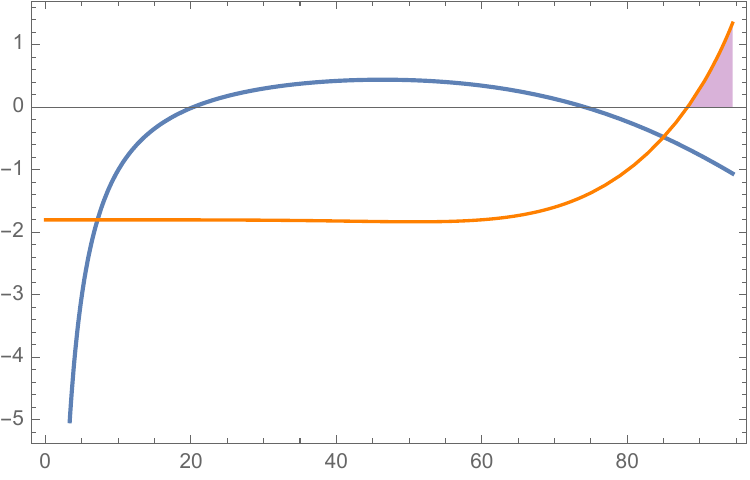} 
    \caption{Plot of $v_s^2$ (orange) and $f(r)$ (blue) as functions of $r$, considering the parameters $\rho_0=0.00003$, $\alpha=2.4$, $\beta=7.5$, $r_v=100$, $r_s=80$, $M=10.0$, and $\delta_c=-0.90$.}
    \label{figstab}
\end{figure}

In this subsection, we discuss the stability of the obtained solution by analyzing the sound speed of the fluid, via
\begin{equation}
    v_s^2=\frac{d\langle p \rangle}{d\rho},
\end{equation}
where $\langle p \rangle = (p_r+2 p_t)/3$ is the average between the radial and lateral pressures. The stability of the solution becomes ensured where $0<v_s^2<1$ \cite{Crispim:2024yjz}. Considering the conservation law of the energy-momentum tensor and the fact that $\rho(r)=-p_r(r)$, valid for the void surrounding the black hole, we have 
\begin{equation}
    \nabla_{\mu}T^{\mu\nu}=0 \Rightarrow p_t(r)=-\left[\frac{r}{2}\frac{d\rho(r)}{dr}+\rho(r)\right].
\end{equation}
In Figure \ref{figstab}, we present the square of the sound speed (orange) and the metric coefficient (blue) as functions of \( r \). The (purple) region highlights where \( 0 < v_s^2 < 1 \), indicating stability. Notably, this stable region extends beyond the cosmological-like horizon at \( r > r_s \), placing it within the wall surrounding the void, where $\rho(r)>\rho_0$, according to Eq. (\ref{dens}). This suggests that the overall structure remains stable, while the instability in the inner region of the void arises either from the collapse of the innermost region toward the center or from the expansion of the outer regions due to a de Sitter-like effect. However, this instability gradually diminishes as one approaches the outer horizon.

\section{Thermodynamics}

In this section, we study the thermodynamical behavior associated with the black hole horizon in a region far away from the edge of the void. In this region, as we previously discussed, space-time has a de Sitter-like behavior which is governed by the void parameters. Accordingly, the thermodynamical properties of the black hole horizon will also be affected by these parameters, and they will directly influence the fate of the black hole.

\subsection{Hawking temperature and evaporation time}

The temperature of a black hole can be derived from its superficial gravity from the standard relation
\begin{equation}
    T_H = \frac{\kappa}{4\pi} = \frac{1}{4\pi}\lim_{r \rightarrow r_h}\sqrt{g^{tt}g^{rr}(g_{rr})'},
\end{equation}
which is the well known Hawking temperature. In our cause, it can be found through
\begin{equation}\label{hawktemp}
    T_H=\frac{f'(r_h)}{4\pi}=T_H^S\left\lbrace 1-\frac{ 
 8 \pi r_h^2 \left[1 + \left(\frac{r_h}{r_v}\right)^{\beta} + \delta_c - \left(\frac{r_h}{r_s}\right)^{\alpha} \delta_c\right] \rho_0}{ 1 + \left(\frac{r_h}{r_v}\right)^{\beta}}\right\rbrace,
\end{equation}
where we have used the black hole mass by setting \( f(r) = 0 \), calculating it as a function of the event horizon radius, \( r_h \), according to Eq. (\ref{BHmass}). In Eq. (\ref{hawktemp}), $T_H^S=1/4\pi r_h$ is the usual Hawking temperature for the Schwarzschild black hole. As usual, a prime indicates derivation with respect to the radial coordinate. 
\begin{figure}[t]
    \centering
    \includegraphics[width=0.6\textwidth]{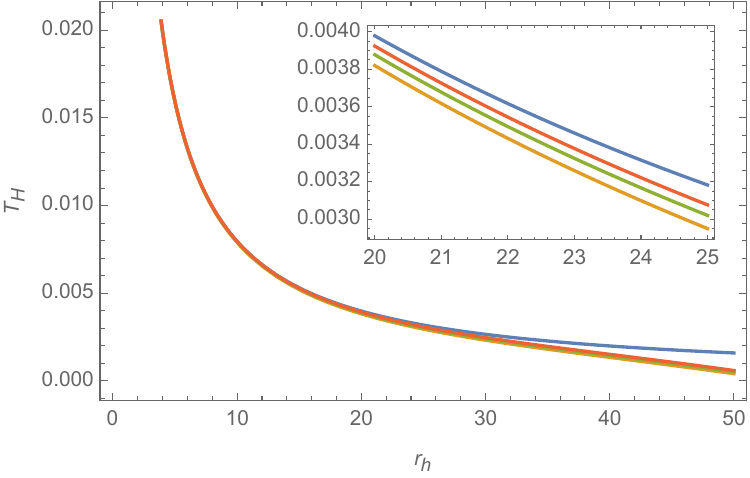} 
    \caption{Hawking temperature $T_H$ as a function of $r_h$, for some values of $\delta_c$, from bottom to top, $\delta_c=-0.90$ (orange), $\delta_c=-0.95$ (green), $\delta_c=-0.99$ (red), and Schwarzschild (blue, $\rho_0=0$), considering the parameters $\rho_0=0.00003$, $\alpha=2.4$, $\beta=7.5$, $r_v=100$, and $r_s=80$.}
    \label{fig2}
\end{figure}

In Fig. \ref{fig2}, we present a plot of the Hawking temperature as a function of the radius of the horizon. The detailed behavior observed in the plot reveals that the temperature increases as the system approaches the absolute vacuum limit. 

The black hole evaporation time is given by
\begin{equation} \label{Estefan}
    \frac{dM}{dt}=\sigma T_H^4 4\pi r_h^2,
\end{equation}
where $\sigma$ is the Stephan-Boltzmann constant (here, we will adopt $\sigma \sim 1$). Thus, the evaporation time for a black hole with size $r_h$ can be integrated via
\begin{equation}
t_e=32 \pi^3 \int_0^{r_h} \frac{u^2 \left[1 + \left(\frac{u}{r_v}\right)^\beta\right]^3}{\left[1 + \left(\frac{u}{r_v}\right)^\beta - 8 \pi u^2 \left(1 + \left(\frac{u}{r_v}\right)^\beta + \delta_c - \left(\frac{u}{r_s}\right)^\alpha \delta_c\right) \rho_0\right]^3} \, du,
\end{equation}
which can only be solved numerically.
\begin{figure}[t]
    \centering
    \includegraphics[width=0.6\textwidth]{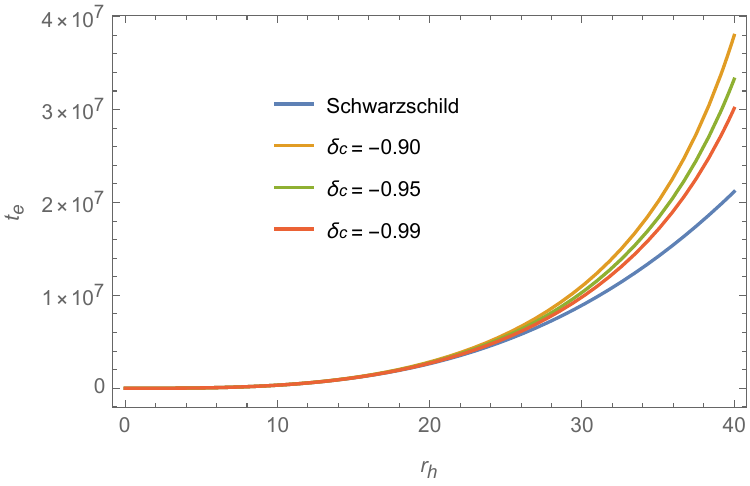} 
    \caption{Black hole evaporation time as a function of $r_h$, for some values of $\delta_c$, considering the parameters $\rho_0=0.00003$, $\alpha=2.4$, $\beta=7.5$, $r_v=100$, and $r_s=80$.}
    \label{fig3}
\end{figure}
In Fig. \ref{fig3} we depict the black hole evaporation time as a function of the radius of the horizon. Note that the greater the modulus of $\delta_c$, the shorter the evaporation time. In other words, as the void becomes more and more emptier, the black hole takes less time to evaporate. Then, in absolute vacuum, when the solution reduces to Schwarzschild, this time is even smaller, becoming the shortest among all cases. The difference between these times is more pronounced for large black holes. These results align with the fact that a higher Hawking temperature corresponds to a shorter evaporation time.

It is well known that the evaporation time of the Schwarzschild solution for large black holes is greater than the age of the universe, which makes it unfeasible at cosmological scales. This is also related to the fact that, given the singular structure of this solution, a complete evaporation would require reaching a limit where the temperature $T \to \infty$ as $r_h \to 0$ \cite{Estrada:2023dcj}. Thus, since the evaporation time in our case study is greater than that of the vacuum case, we can also infer that evaporation would be unfeasible at cosmological scales. The fact that the evaporation time in our case study is greater than that of the vacuum case could be related to the fact that, from equation \eqref{Estefan}, $dM(r_h)/r_h^2 \cdot (T_h^4)^{-1} \sim dt$, since the temperature in our case study is lower than that of Schwarzschild.

\subsection{Entropy and specific heat}

\begin{figure}[t]
    \centering
    \includegraphics[width=0.6\textwidth]{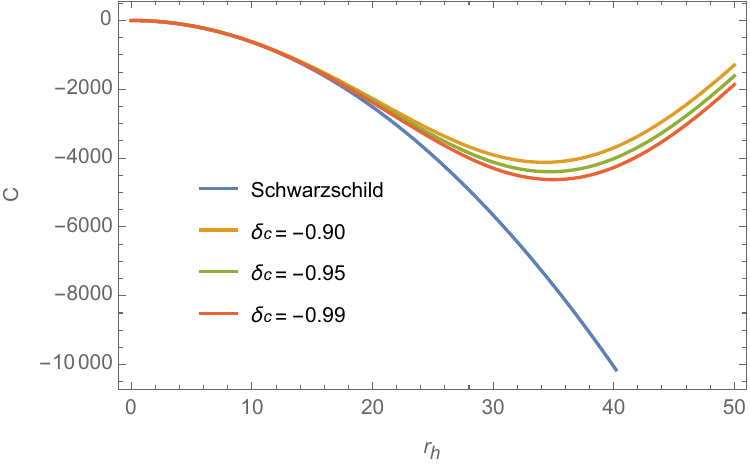} 
    \caption{Black hole specific heat as a function of $r_h$, for some values of $\delta_c$, considering the parameters $\rho_0=0.00003$, $\alpha=2.4$, $\beta=7.5$, $r_v=100$, and $r_s=80$.}
    \label{figC}
\end{figure}

The analysis of the entropy, given by
\begin{equation}
S=\int\frac{dM}{T_H},
\end{equation}
reveals that our black hole solution follows Bekenstein's area law, $S=\pi r_h^2$. On the other hand, the specific heat, $C$, expressed as
\begin{equation}
    C=\frac{dM}{dT_H}=\frac{dM/dr_h}{dT_H/dr_h},
\end{equation}
yields an involved expression that significantly deviates from the Schwarzschild solution for large black holes. In Figure \ref{figC}, we can see that, in a cosmic void, the black hole's specific heat reaches a minimum, signaling a shift in its thermal behavior. Below this point, evaporation accelerates, while beyond it, energy loss slows. This suggests that the void tends to stabilize the black hole, regulating its evolution, with the presence of a de Sitter-like horizon moderating the black hole evaporation. Although the analysis of this section is restricted to the event horizon associated with the black hole, we can see from the behavior of the specific heat that our solution signals a different thermodynamical evolution when the event horizon is sufficiently large. In this case, we can observe that the black hole-void system reduces the thermodynamic instability due to the nature of the external de Sitter-like horizon.

\section{A glimpse into the interaction between the event horizon and the cosmological horizon.}

In the analysis carried out above, the studied scenario was where the event horizon contracts and evaporates in a finite time. Therefore, in this analysis, the potential influence of the cosmological horizon on the thermodynamic evolution of the event horizon was not considered. However, as we can see in Figure \ref{figC}, from left to right, the specific heat begins to increase after reaching a minimum, which suggests that as the event horizon grows, coinciding with a decrease in its temperature, the black hole receives more energy from its exterior. Thus, this fact could suggest that the increase in energy might be due to interaction with the cosmological horizon. In other words, the event horizon could be receiving more energy as it approaches the cosmological horizon. 

In this way, in this section, we will analyze the opposite case to the previous one, i.e., instead of the event horizon contracting and evaporating, we will study the possibility of the event horizon expanding while thermally interacting with the cosmological horizon. In the literature, the hypothesis that the cosmological horizon may also emit a temperature, leading to a thermodynamic interaction between both horizons, has already been studied in other different scenarios. See, for example, references \cite{Aros:2008ef,Kubiznak:2015bya, Qiu:2019qgp,Estrada:2024uuu}. We will shed light on the interaction between both horizons; however, a more in-depth theoretical study is still required in future work. 

First, we can observe the following in the parameter space using the triple product chain rule, where $f(r = r_*) = 0$. Thus, $r_*$ can correspond to either the event horizon or the outer cosmological horizon:

\begin{equation} \label{TripleChainRule}
    \delta f =0= \frac{\partial f}{\partial r_*} dr_* + \frac{\partial f}{\partial M} dM \rightarrow \frac{\partial f}{\partial r_*} \frac{\partial r_*}{\partial M} \frac{\partial M}{\partial f} =-1 \rightarrow T_* \sim \frac{\partial f}{\partial r_*} \sim -\frac{\partial f}{\partial M} \frac{\partial M}{\partial r_*} \sim \frac{1}{r_*} \frac{\partial M}{\partial r_*}
\end{equation}

By definition, the derivative $\frac{\partial M}{\partial r_*}$ is positive at the event horizon, i.e., $\frac{\partial M}{\partial r_h} > 0$, see figure \ref{figMasa1}, thus: 

\begin{equation}
    T_h \sim \frac{\partial f}{\partial r_h} \sim \frac{1}{r_h} \frac{\partial M}{\partial r_h}>0 \Rightarrow dU=dM=T_h r_h dr_h
\end{equation}
and the heat capacity at the event horizon is given by:
\begin{equation}
    C_h=dU/dT_h=dM/dT_h.
\end{equation}

The mentioned derivative is negative at the cosmological horizon, i.e., $\frac{\partial M}{\partial r_{++}} < 0$, see figure \ref{figMasa1}. Additionally, by definition, the surface gravity is such that at the cosmological horizon, the temperature is positive \cite{Kubiznak:2015bya,Qiu:2019qgp}. Thus,
\begin{equation}
    T_{++} \sim - \frac{\partial f}{\partial r_{++}} \sim - \frac{1}{r_{++}} \frac{\partial M}{\partial r_{++}}>0 \Rightarrow dU=-dM=T_{++} r_{++} dr_{++}
\end{equation}
and the heat capacity at the  cosmological horizon:
\begin{equation}
    C_{++}=dU/dT_h=-dM/dT_{++}
\end{equation}
The mentioned derivatives vanish at the critical case where both horizons coincides, i.e., $\frac{\partial M}{\partial r_{cri}} =0$, see Figure \ref{figMasa1}. Thus
\begin{equation}
    T_{cri} \sim  \frac{\partial M}{\partial r_{cri}}=0.
\end{equation}
consequently $C=0$ at this critical point. 

\begin{figure}[ht]
    \centering
    \includegraphics[width=0.6\textwidth]{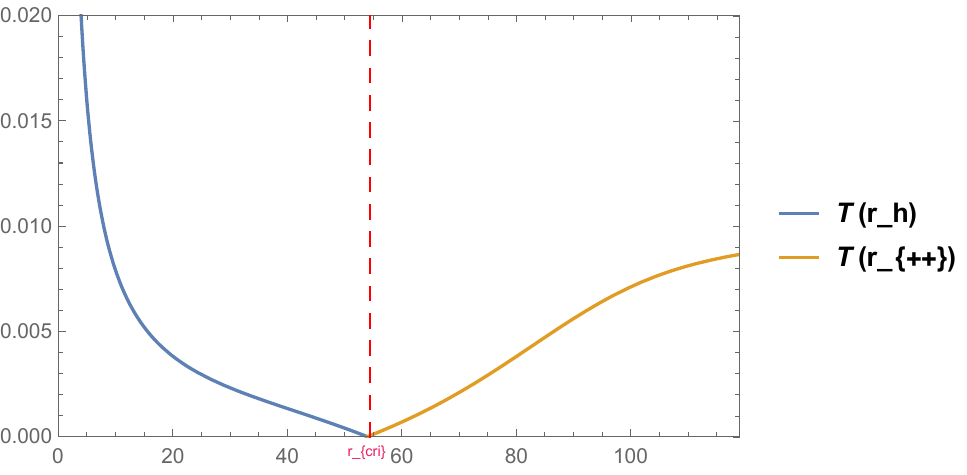} 
    \caption{The horizontal axis corresponds to the $r$ such that $f(r)=0$. At the left curve corresponds to $r_h$. At the right curve corresponds to the $r_{++}$. $r_s=80,r_v=100, \delta=-0.9,\rho_0=0.00003$}
    \label{figTemperaturaAmbos}
\end{figure}

From the previous analysis, we can display the temperature of both horizons in Figure \ref{figTemperaturaAmbos}. Remarkably, we can notice that at the value where both horizons coincide, $r_{\text{cri}} = r_h = r_{++}$, the temperature vanishes. In Figures \ref{figMasa1} and \ref{figTemperaturaAmbos}, this occurs for $r_{\text{cri}} \sim 54.5$. Consequently, the temperature also vanishes for $M_{\text{cri}} = M(r_{\text{cri}})$.

Since the mass parameter $M$ is the same for each ordered pair $(r_h, r_{++})$, following references \cite{Aros:2008ef,Estrada:2019qsu,Estrada:2024uuu}, we will also assume that its variation is the same. Thus, from the previous equations, we can write:
\begin{equation}
    dM=T_h r_h dr_h=-T_{++} r_{++} dr_{++}
\end{equation}
 Thus, the radial evolution of both horizons is interconnected. For example, if $dM > 0$, the event horizon expands with $dr_h > 0$ and the cosmological horizon contracts with $dr_{++} < 0$, and vice versa. On the other hand, if $dM = 0$, both horizons would remain static. We will now analyze this in detail:

 \begin{itemize}
     \item In the ascending curve of Figure \ref{figMasa1}, from left to right, we observe that the mass parameter increases at the event horizon, i.e., $dM > 0$. From Figure \ref{figTemperaturaAmbos}, we observe that, also from left to right, the temperature decreases. In this way, the heat capacity at the event horizon is $C_h = dU/dT_h = dM/dT_h < 0$. That is, the event horizon is receiving radiation energy from the cosmological horizon as its temperature decreases. In this process, the event horizon expands.
     \item In the descending curve of Figure \ref{figMasa1}, from right to left, we observe that the mass parameter increases at the cosmological horizon, i.e., $dM > 0$. From Figure \ref{figTemperaturaAmbos}, we observe that, also from right to left, the temperature decreases. In this way, the heat capacity at the cosmological horizon is $C_{++} = dU/dT_{++} = -dM/dT_{++} > 0$. That is, the cosmological horizon is releasing radiation energy toward the event horizon as its temperature decreases. In this process, the cosmological horizon contracts.
     \item While the event horizon expands, increasing its mass and decreasing its temperature, the cosmological horizon contracts, also increasing its mass and decreasing its temperature. This process will continue as long as the specific heat of the event horizon is negative, and the specific heat of the cosmological horizon is positive. That is, the cosmological horizon releases energy, which is absorbed by the event horizon. However, at the peak of Figure \ref{figMasa1}, which coincides with the point where the temperature vanishes in Figure \ref{figTemperaturaAmbos}, the specific heat of both horizons also vanishes.At this point, where $C_h = C_{++} = 0$, i.e., the cosmological horizon stops releasing energy towards the event horizon (and consequently the event horizon stops absorbing energy), the radial evolution of both horizons ceases, leaving a remnant whose mass corresponds to the critical mass and whose radius corresponds to the value where both horizons coincide, $r_{\text{cri}} = r_h = r_{++}$.
 \end{itemize}

In this way, we have shed light on the interaction between both horizons; however, a more in-depth theoretical study should be required in future work to, for example, propose a formalism to determine the possibility of this interaction occurring in a finite time.

\section{Discussion and Summary}
In this work, we investigated the properties of static and spherical black holes that occupy the center of cosmic voids, deriving analytically a new metric that accounts for the influence of the universal density profile associated with the void, presented in \cite{Hamaus_2014_2}, which we supposed to be a source for these black holes. Our analysis revealed that the presence of the void significantly modifies the black hole structure, leading to an additional horizon besides the event horizon. This second horizon, analogous to a cosmological horizon, emerges due to the de Sitter-like nature of the void at large distances. We found that as the absolute value of the contrast parameter $\delta_c$ increases, the size of the black hole horizon increases, while that of the cosmological horizon decreases. Furthermore, in all cases, the inner horizon remains larger than the Schwarzschild radius, reinforcing the impact of the void on the gravitational structure. 

Thus, our analysis revealed that the cosmic void surrounding the black hole significantly modifies its properties. Vice versa, our findings seem also to indicate that the black hole contributes to the void stability, since there exists a region between the singularity and the edge of the void whose spacetime resembles a de Sitter one, introducing a repulsive effect that counteracts possible gravitational collapses and stabilizes the system. In this sense, the cosmic void hosting the black hole must expand faster than the overall cosmic expansion, since the void effective cosmological constant and the background one must be mutually reinforced. In a more speculative direction, in a universe with no cosmological constant a sufficient number of black holes in voids could account for the accelerated expansion we observe today, in line with recent discussions in the literature \cite{Seifert2024, Williams2024}.

{We also examined the stability of the obtained solution by computing the sound speed of the fluid associated with the void and surrounding the black hole. Our results confirm the presence of a stable region within the void’s wall at \( r > r_s \), near the cosmological-like horizon of the black hole. This stability helps maintain the overall structure, except for the inner region, which continues to evacuate, further deepening the void. The instability within the void arises either because the innermost regions tend to collapse toward the center or because the outer regions tend to expand outward due to the dS-like behavior.}

Examining the black hole’s thermodynamics, we found that its Hawking temperature diminishes as the void deviates of a perfect (classical) vacuum, leading to a greater evaporation time compared to the standard Schwarzschild case, notably for large black holes. In addition, the specific heat exhibits a minimum, signaling a transition in the thermal behavior of the black hole, where evaporation initially accelerates but then slows down because of the influence of the void. Thus, such thermal effects are more pronounced in deeper voids, {\it i.e.}, those with a higher absolute contrast parameter.

The behavior of the specific heat at the event horizon, such that, being negative, it increases from left to right, tending to reach zero, suggests that this could be due to a potential interaction with the cosmological horizon. In other words, the event horizon may be absorbing more energy from the cosmological horizon as the distance between them decreases. In this way, we have provided a glimpse into this potential interaction. Specifically: While the event horizon expands, increasing its mass and decreasing its temperature (i.e., its specific heat is negative), the cosmological horizon contracts, also increasing its mass and decreasing its temperature (i.e., its specific heat is positive). This process will continue as long as the specific heat of the event horizon is negative, and the specific heat of the cosmological horizon is positive. That is, the cosmological horizon releases energy, which is absorbed by the event horizon. However, at the peak of Figure \ref{figMasa1}, which coincides with the point where the temperature vanishes in Figure \ref{figTemperaturaAmbos}, the specific heat of both horizons also vanishes. At this point, where $C_h = C_{++} = 0$, i.e., the cosmological horizon stops releasing energy towards the event horizon (and consequently the event horizon stops absorbing energy), the radial evolution of both horizons ceases, leaving a remnant whose mass corresponds to the critical mass and whose radius corresponds to the value where both horizons coincide, $r_{\text{cri}} = r_h = r_{++}$.

Our model serves as a starting point for further investigation into the possible effects of black holes inside voids on the overall cosmological evolution of the universe. Although it has generally been assumed that black holes can be ignored in cosmic evolution, recent works have proposed both phenomenological and theoretical arguments suggesting a potential connection between the two \cite{Cadoni_2023, Faraoni2024}. On the other hand, models incorporating cosmic voids have been extensively studied numerically, indicating the need for a better understanding of their impact on observations of standard candles \cite{Seifert2024, Williams2024}. We discussed how the presence of black holes could lead to the formation de Sitter-like bubbles, though a complete study of these possibilities would also require the inclusion of dynamical effects beyond the scope of our analysis. Nevertheless, our results highlight the potential of studying more refined models of black holes within cosmic voids to enhance our understanding of cosmic evolution.

\section*{Acknowledgements}
FBL is funded by Fundação Cearense de Apoio ao Desenvolvimento Científico e Tecnológico (FUNCAP) and by  Conselho Nacional de Desenvolvimento Científico e Tecnológico (CNPq), grant number 305947/2024-9. ME is funded by ANID, FONDECYT de Iniciaci\'on en Investigación 2023, Folio 11230247. CRM and MSC are partially funded by Conselho Nacional de Desenvolvimento Científico e Tecnológico (CNPq), under the grants 308268/2021-6 and 315926/2021-0, respectively.  
\bibliographystyle{JHEP}
\bibliography{ref.bib}
\end{document}